\documentstyle[epsfig,12pt]{article}
\def\Tr{\mathop{{\cal T}\mskip-4.5mu \lower.1ex \hbox{\it r}\mskip+4.5mu}}
\begin{document}
\pagestyle{empty}
\rightline{\vbox{
\halign{&#\hfil\cr
& OCIP/C-95-8 \cr
& TTP95-24 \cr
& hep-ph/9507220 \cr
& July 1995 \cr}}}
\bigskip
{\Large\bf
\centerline{Fragmentation production of doubly heavy baryons}}
\bigskip
\normalsize
\bigskip
\centerline{Michael A. Doncheski}
\centerline{\sl Ottawa-Carleton Institute for Physics, Department of
          Physics,}
          \centerline{\sl Carleton University, Ottawa, Ontario K1S 5B6,
          Canada}
\bigskip
\centerline{and}
\bigskip
\centerline{J.\ Steegborn and M.\ L.\ Stong}
\centerline{\sl Institut f\"ur Theoretische Teilchenphysik, Universit\"at, }
\centerline{\sl D-76128 Karlsruhe, Germany}
\bigskip
\bigskip
\begin{abstract}
Baryons with a single heavy quark are being studied experimentally at
present.  Baryons with two units of heavy flavor will be abundantly produced
not only at future colliders, but also at existing facilities.  In this
paper we study the production via heavy quark fragmentation of baryons
containing two heavy quarks at the Tevatron, the LHC, HERA, and the NLC.
The production rate is woefully small at HERA and at the NLC, but
significant at $pp$ and $p\bar{p}$ machines.  We present distributions in
various kinematical variables in addition to the integrated cross sections
at hadron colliders.
\end{abstract}
\newpage
\pagestyle{plain}\setcounter{page}{1}

The constituent quark model has been remarkably successful in describing the
observed hadronic states.  This includes mesons with one heavy and one light
quark ({\it i.e.}, $B$ and $D$ mesons), baryons with one heavy and two light
quarks, and the $J/\psi$ and $\Upsilon$ mesons, which contain a heavy quark
and heavy antiquark.  Recent experiments have made great progress in the
observations of the $c$ and $b$ baryons\cite{bar_a,bar_c} and their
decays\cite{bar_dec_a,bar_dec_c}.  The constituent quark model predicts the
existence of baryons containing two heavy quarks ($cc$, $bc$, or $bb$) and
one light quark, and we are on the verge of obtaining the necessary
experimental sensitivity to observe these states.  The energies necessary to
produce these particles are already reached; the difficulty remaining is in
their reconstruction.  These states have in general a large number of decay
modes, so that their observation and a measurement of their properties will
require a large number of them to be produced.  This difficulty is increased
by the fact that the production rates at $e^+ e^-$ colliders are extremely
small, so that the identification of these particles must take place in the
messier environment of hadronic collisions.

Were the heavy quark of infinite mass, the two heavy quarks would be bound
in a point-like diquark, and the light degrees of freedom would ``see'' this
diquark as a static antitriplet color source.  In this limit, the
interactions of the light degrees of freedom with the heavy diquark are
quite similar to interactions of the light degrees of freedom with the
$\bar{b}$ quark in a $B$ meson\cite{savagewise,whitesavage}.  For realistic
heavy quark masses, this simple picture is not yet valid.  In particular,
the hyperfine splittings of these baryons are not yet well described by the
heavy-mass limit\cite{ml}.  The states may nonetheless be accurately treated
as a combination of light degrees of freedom and heavy although
not-pointlike diquark.  The interactions between the two heavy quarks are
analogous to the $Q\overline{Q}$ system familiar from $J/\psi$ and
$\Upsilon$ spectroscopy.  At short distances, the interaction will be
dominated by one-gluon-exchange, although with a color factor 2/3 from the
$QQ$ interaction rather than 4/3 from $Q\overline{Q}$, and at long distances
it will be confining\cite{fleck}.

The production of these states is reliably calculable as a hard process.
One takes the full set of Feynman diagrams for the production of two pairs
of heavy quarks and heavy antiquarks and requires that the momenta of the
two heavy quarks be nearly equal, then projects out the correct quantum
numbers for the relevant $Q_1Q_2$ state.  This is similar to the calculation
of the hard production of heavy quarkonium states\cite{baierruckl}, though
more complicated as an additional $Q\overline{Q}$ pair is needed.  On the
other hand, because the quarks are heavy, fragmentation functions for a
heavy quark to produce a doubly heavy diquark can be predicted in
perturbative QCD\cite{falk,qfrag}.  The fragmentation description of the
$Q_1Q_2$ production process is not valid where the masses of the heavy
quarks become important, that is, for small $p_{_T}$ or small energies, but
for high-energy colliders such as LEP or Tevatron it should be applicable.
The calculation is similar to that described in \cite{qfrag} for $J/\psi$
production.  The form of the matrix element is different in the two cases,
requiring a somewhat changed method of projection onto the proper spin
states.  For the production of a $Q\overline{Q}$ bound state, the standard
procedure is to replace the heavy quark production matrix element
$\bar{u}{\cal O}v = \Tr ({\cal O}v\bar{u})$ with the trace
$\Tr ({\cal O}P_{SS_z})$, where $P_{SS_z}$ is the proper projection onto the
spin of the $Q\overline{Q}$ state\cite{kuehn}.  This procedure is, however,
not simply applicable to the $Q_1Q_2$ bound states due to the form of the
matrix element, which contains not $u$ and $\bar{v}$ as in the production of
$Q\overline{Q}$ but $u_1$ and $u_2$.  The projection procedure described
above is nothing more than a clever method of performing the sum over the
quark spins.  This sum can also be performed by summing over helicity
amplitudes with the proper Clebsch-Gordan coefficients.  Although the
procedure is somewhat different, the calculations are identical in their
essentials, and the fragmentation function is simply proportional to that
for a $Q\overline{Q}$ pair.  The differences reflect the changed color and
statistical factors.  The fragmentation function for $Q_1$ to produce a
spin-1 diquark $(Q_1Q_2)$ is
\begin{eqnarray}
& &D_{Q_1\to(Q_1Q_2)}(z,\mu=(m_1+2m_2)) =
{2N_{12} \over 9 \pi}{|R_{(Q_1Q_2)}(0)|^2 \over m_2^3}
\alpha_s^2(2m_2) F(z), \label{eq:D_lowmom} \\
& &F(z) =  {rz (1-z)^2 \over 4(1 - z + rz)^6}
\left[ 2 - 2z(3-2r) + 3z^2(3-2r+4r^2) \right. \nonumber \\
&& \hphantom{F(z) = dum} \left.
-2z^3(1-r)(4-r+2r^2) - z^4(1-r)^2(3-2r+2r^2) \right], \label{eq:F_spin1}
\end{eqnarray}
where $r = m_2/(m_1+m_2)$.  The factor $N_{12} = (1)2$ for (un)equal quark
masses, and reflects the presence of identical fermions in the final state.
For production of a spin-0 state, $(Q_1Q_2)^\prime$, the function $F(z)$
must be replaced by
\begin{eqnarray}
F^\prime(z) & = & {rz (1-z)^2 \over 12 (1 - z + rz)^6}\left[ 6 - 18 z(1-2 r)
+ z^2(21-74r+68r^2) \right. \nonumber \\ && \hphantom{dum}\left.
- 2z^3(1-r)(6-19r+18r^2) + 3z^4(1-r)^2(1-2r+2r^2) \right]. \label{eq:F_spin0}
\end{eqnarray}
As in the case of hard quarkonium production, the requirement of nearly
equal momenta and the projection onto the proper $Q_1Q_2$ quantum numbers
will modify the $p_{_T}$ spectrum of the diquark; it will fall more rapidly
than that of single heavy quark production.  For heavy quarkonium
production, the faster fall-off in $p_{_T}$ of the hard production process
means that fragmentation production will dominate for sufficiently large
$p_{_T}$ despite the suppression of the fragmentation production by powers
of $\alpha_s$\cite{us1}.  For the $Q_1Q_2$ states, no hard production
processes of lower order in $\alpha_s$ exist.  We assume that the doubly
heavy diquarks will always hadronize into baryons with two heavy quarks, so
that the fragmentation function
$D_{Q_1 \to {\rm baryon}} = D_{Q_1 \to (Q_1Q_2)}$.

A $Q_1Q_2$ diquark in the ground state has a symmetric spatial
wavefunction.  Because the $Q_1Q_2$ must combine with a light quark to
produce a colorless baryon, the diquark must be a color antitriplet
(antisymmetric) state.  For identical quarks the wavefunction must satisfy
Fermi-Dirac statistics, and thus the $cc$ and $bb$ ground state diquarks
have spin 1.  The $bc$ states are not restricted by statistics, and we
denote the spin singlet (triplet) state by $bc^\prime$ ($bc$).  The spin
triplet can hadronize into a spin-1/2 or -3/2 baryon, $\Xi_{bc}$ and
$\Xi_{bc}^*$, respectively; the spin singlet can produce only the spin-1/2
$\Xi'_{bc}$.  The probabilities for the fragmentation of heavy quarks into
these baryons are: $c \to \Xi_{cc},\Xi^*_{cc}$, about $2 \times 10^{-5}$;
$b \to \Xi'_{bc}$, about $4 \times 10^{-5}$; $b \to \Xi_{bc}, \Xi^*_{bc}$,
about $5 \times 10^{-5}$.  The remaining probabilities are suppressed by
$\sim (m_c/m_b)^3$ and therefore approximately two orders of magnitude
smaller\cite{falk}.

In addition to these probabilities, it is desirable to have predictions of
the $p_{_T}$ (or other kinematical variables) distributions for the
production of these states.  The fragmentation functions given above at low
momentum scale $\mu_0$ must be evolved to higher scales $\mu$ in order to be
of use in calculations for high energy colliders:
\begin{equation}
d\sigma(A + B \to (Q_1Q_2) + X) = \sum_i \int_0^1 dz \;
d\sigma(A + B \to i(\frac{p_{_T}}{z},\eta)+X, \mu^2) \;
D_{i\to(Q_1Q_2)}(z,\mu^2).
\label{sig}
\end{equation}
The evolution occurs via an Altarelli-Parisi equation:
\begin{equation}
\mu^2 \frac{\partial}{\partial \mu^2} D_{i \to (Q_1Q_2)}(z, \mu^2) =
\frac{\alpha_s}{2\pi}
\sum_{j} \int^{1}_{z} \frac{dy}{y} \; P_{ij}(z/y)
\; D_{j \to (Q_1Q_2) }(y, \mu^2).
\label{ap}
\end{equation}
which does not change the overall probabilities given above.  In the
evolution of the fragmentation functions via (\ref{ap}), we include only the
$P_{QQ}$ splitting function.  The fragmentation of gluons to heavy diquarks
is suppressed relative to that of heavy quarks by $\alpha_s$, and may be
neglected.  A separate numerical evolution of the fragmentation functions to
each $z$ and $\mu$ needed in the course of a numerical simulation would be
an extremely computer-intensive approach; instead we generate values of the
fragmentation function for a reasonably-spaced set of $\mu$ and $z$ values
and interpolate to the desired points.  The unevolved fragmentation
functions depend on $|R(0)|^2$, the non-relativistic radial wavefunction at
the origin, the heavy quark masses, and $\alpha_s(\mu_0)$.  These input
parameters are given in Table~\ref{tab:frag_par}.

\begin{table}[h]
\begin{center}
\begin{tabular}{lccc}
\hline \hline
                    &  $\alpha_s$ & $m_Q$   &    $|R(0)|^2$  \\ \hline
$c \to (cc)$        & 0.238       & 1.5 GeV & (0.65 GeV)$^3$ \\
$c \to (bc)$        & 0.173       & 4.5 GeV & (0.80 GeV)$^3$ \\
$c \to (bc)^\prime$ & 0.173       & 4.5 GeV & (0.80 GeV)$^3$ \\
$b \to (bb)$        & 0.173       & 4.5 GeV & (1.20 GeV)$^3$ \\
$b \to (bc)$        & 0.238       & 1.5 GeV & (0.80 GeV)$^3$ \\
$b \to (bc)^\prime$ & 0.238       & 1.5 GeV & (0.80 GeV)$^3$ \\ \hline \hline
\end{tabular}
\end{center}
\caption{Parameters used to calculate the unevolved fragmentation functions}
\label{tab:frag_par}
\end{table}

The $J/\psi$ and $\Upsilon$ decays to lepton pairs give the wavefunctions
$|R(0)|^2$ for these states relatively accurately\cite{schuler} and
independently of phenomenological potential models.  The QCD-corrected
expression for the leptonic decay of heavy quarkonium states
\begin{equation}
\Gamma(V \to \ell^+ \ell^-) = \frac{4 \alpha^2 e_Q^2}{M_V^2} |R_V(0)|^2
\left[ 1 - \frac{16}{3 \pi} \alpha_s(m_Q) \right]
\label{wfn_data}
\end{equation}
is used to extract the $\Upsilon$ and $J/\psi$ radial wavefunction at the
origin.  There is no information on the $B_c$ meson to determine the
wavefunction for the $b\bar{c}$ states, and the annihilation of this state
into lepton pairs would not exist in any case to provide the measurement.
However, there is an empirical relationship among leptonic decays of the
$\phi$, $J/\psi$ and $\Upsilon$ mesons,
$\Gamma(V \to e^+e^-) \simeq 12 e_Q^2 \; {\rm keV}$, which indicates that
$|R(0)|^2/\mu^2$ is approximately constant.

The one-gluon exchange contribution for $Q_1Q_2$ differs from that for
$Q_1\overline{Q}_2$ by a relative color factor 1/2.  In a Coulomb potential,
therefore, the relation $|R_{J/\psi}(0)|^2 = 8 |R_{cc}(0)|^2$ must hold.
The heavy quarks are, however, not in a pure Coulomb potential, so that we
must rely on phenomenological potentials to better estimate the radial
wavefunction.  While many functional forms for the potential provide quite
good fits to the heavy quarkonium data, the fitted potentials all have very
similar shapes in the region of $r \sim 1 \; {\rm GeV}^{-1}$.  We use 1/2
the $Q_1\overline{Q}_2$ potential for the $Q_1Q_2$ potential, and find the
wavefunction at the origin using a numerical solution of the Schr\"odinger
equation.  We compare in Table~\ref{tab:wfn_orig} the results of this
calculation for two familiar potentials with the values extracted from the
$J/\psi$ and $\Upsilon$ decay rates.

\begin{table}[h]
\begin{center}
\begin{tabular}{rccc}\hline \hline
& Indiana\cite{indiana} & Richardson\cite{richardson} & data\cite{pdg}\\
\hline
$c\bar{c}$ & (0.95 GeV)$^3$ & (0.93 GeV)$^3$ & (0.98 GeV)$^3$ \\
$cc$       & (0.67 GeV)$^3$ & (0.65 GeV)$^3$ &                \\
$b\bar{c}$ & (1.18 GeV)$^3$ & (1.18 GeV)$^3$ &                \\
$bc$       & (0.82 GeV)$^3$ & (0.81 GeV)$^3$ &                \\
$b\bar{b}$ & (1.72 GeV)$^3$ & (1.88 GeV)$^3$ & (1.96 GeV)$^3$ \\
$bb$       & (1.15 GeV)$^3$ & (1.23 GeV)$^3$ &                \\ \hline
\end{tabular}
\end{center}
\caption{Radial wavefunctions at the origin, $|R(0)|^2$, for
$Q_1\overline{Q}_2$ and $Q_1Q_2$ states for two interquark potentials
compared to values extracted from data, see (6).}
%Eqn.~\ref{wfn_data}.}
\label{tab:wfn_orig}
\end{table}

Doubly heavy baryons will be produced at all present and future accelerators
where there is sufficient energy.  The results presented here include both
baryon and antibaryon production.  We consider first the $e^+e^-$ colliders
LEP, LEPII, and NLC.  At LEP, the only subprocess for the production of a
heavy quark $Q$ is $e^+e^- \to Z$, with the $Z$ decaying to
$Q\overline{Q}$.  The production rates in this case are quite small, a few
events per year.  For LEPII and NLC the additional processes
$e^+e^- \to W^+W^-$ where one $W$ decays to a $c$ quark and $e^+e^- \to ZZ$
where one $Z$ decays to $c\bar{c}$ or $b\bar{b}$ also contribute (depending
on the LEPII energy, of course).  Despite the additional subprocesses, of
which $W^+W^-$ dominates for NLC, the event rates remain woefully small.  At
LEPII, only a few events/year can be expected, and a 60~fb$^{-1}$/yr,
500~GeV linear collider will produce only of order 30 events per year.  The
production rates at $\gamma\gamma$ colliders and $e\gamma$ colliders were
also calculated, including backscattered and Weizs\"acker-Williams photons
and the contributions from resolved-photon processes.  The cross sections
for production of these particles at these colliders are also hopelessly
small.

The results at HERA are similar.  The main subprocesses here are
$\gamma g \to Q\overline{Q}$ and $\gamma Q \to gQ$.  Assuming the design
luminosity of 200~pb$^{-1}$/yr, we again find that the event rate is rather
small, of order 30 events/year.  The resolved photon processes are likewise
negligible.

The situation is considerably more hopeful at hadron colliders.  The
relevant subprocesses are $gg \to Q\overline{Q}$,
$q\bar{q} \to Q\overline{Q}$, $gQ \to gQ$, and $qQ \to qQ$ (in the latter
two subprocesses, $q$ and $Q$ stand for both quarks and antiquarks),
although the main contribution is from the gluon-fusion subprocess.  For the
Tevatron, with an integrated luminosity of 100~pb$^{-1}$/yr, we anticipate
of order $8\times10^4$ events/year ($p_{_T} > 5$~GeV, $|\eta| < 0.5$), while
the LHC, with 100~fb$^{-1}$/yr, should produce of order $1.3\times10^8$
events per year($p_{_T} > 10$ GeV, $|\eta| < 0.5$).  Total production cross
sections per unit rapidity for the various diquarks at the Tevatron and the
LHC (operating at both 10~TeV and 14~TeV) are given in
Table~\ref{tab:tot_sig}.  The rates for production of the $bc$-type states
are similar to those expected for the $B_c$ meson, where rates of 10$^4$
$B_c$'s have been predicted at the Tevatron ($p_{_T}>10$~GeV) and 10$^7$ at
LHC ($p_{_T}>20$~GeV)\cite{kol2}.  On the other hand, the production rate
for $cc$-type states is significantly smaller than that for $J/\psi$
production at the Tevatron which is expected to be $5\times 10^6$
($p_{_T}>5$~GeV)\cite{us1}.

\begin{table}[h]
\begin{center}
\begin{tabular}{lrrr}
\hline \hline
                         & Tevatron & LHC (10 TeV) & LHC (14 TeV) \\ \hline
 $\Xi_{cc} + \Xi_{cc}^*$ & 430 pb & 330 pb & 470 pb \\
 $\Xi_{bc}^\prime$       & 145 pb & 220 pb & 330 pb \\
 $\Xi_{bc} + \Xi_{bc}^*$ & 215 pb & 350 pb & 490 pb \\
 $\Xi_{bb} + \Xi_{bb}^*$ &  16 pb & 27 pb  &  36 pb \\   \hline
\end{tabular}
\end{center}
\caption{Production cross sections per unit rapidity in the central region.}
\label{tab:tot_sig}
\end{table}

\begin{figure}[h]
\begin{minipage}{0.49\linewidth}
\epsfig{file=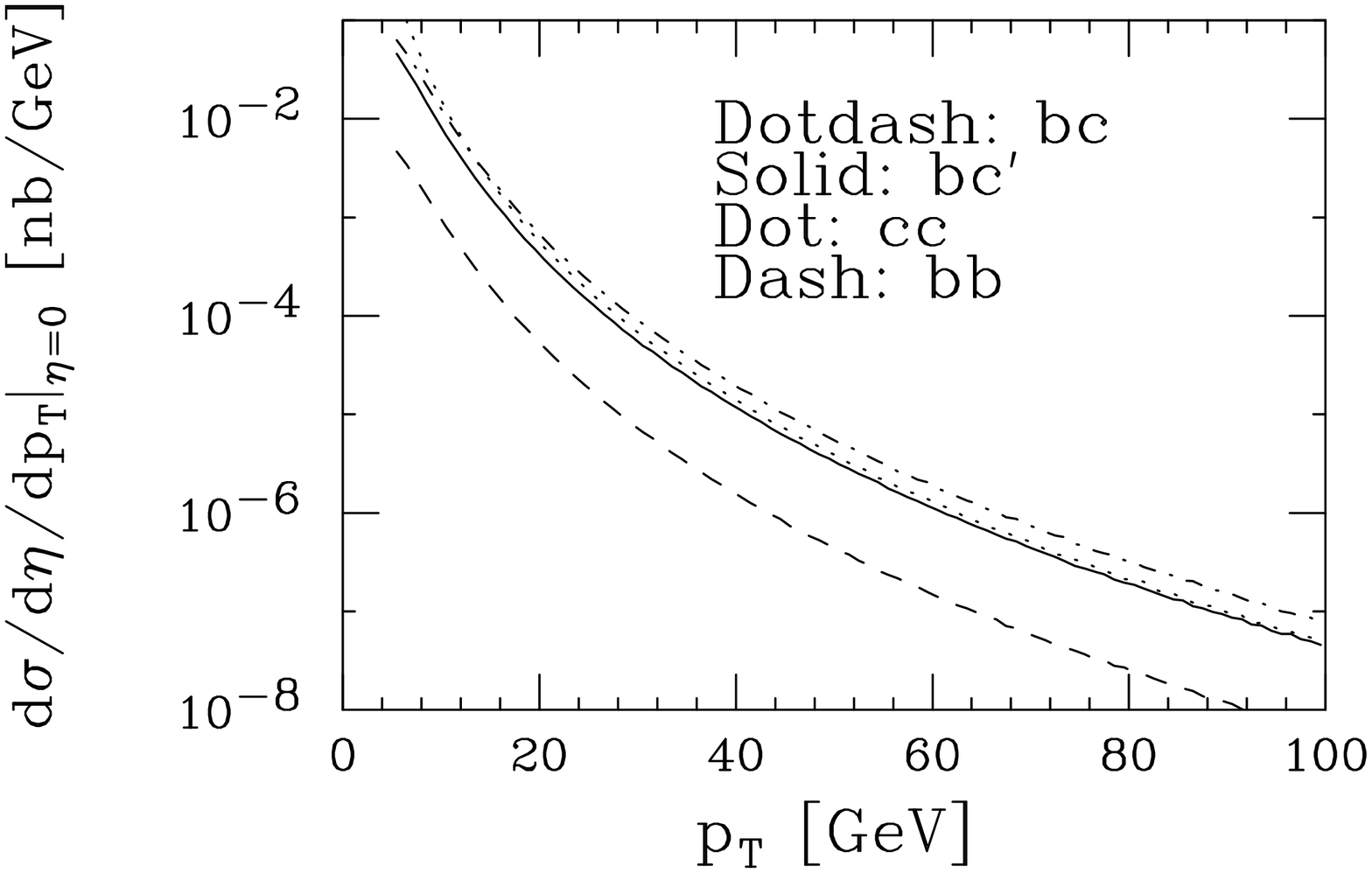,width=\linewidth}
\caption{$p_{_T}$ distributions (at small rapidity) for doubly heavy baryon
production at the Tevatron.}
\label{fig:pt_tev}
\end{minipage}\hfill
\begin{minipage}{0.49\linewidth}
\epsfig{file=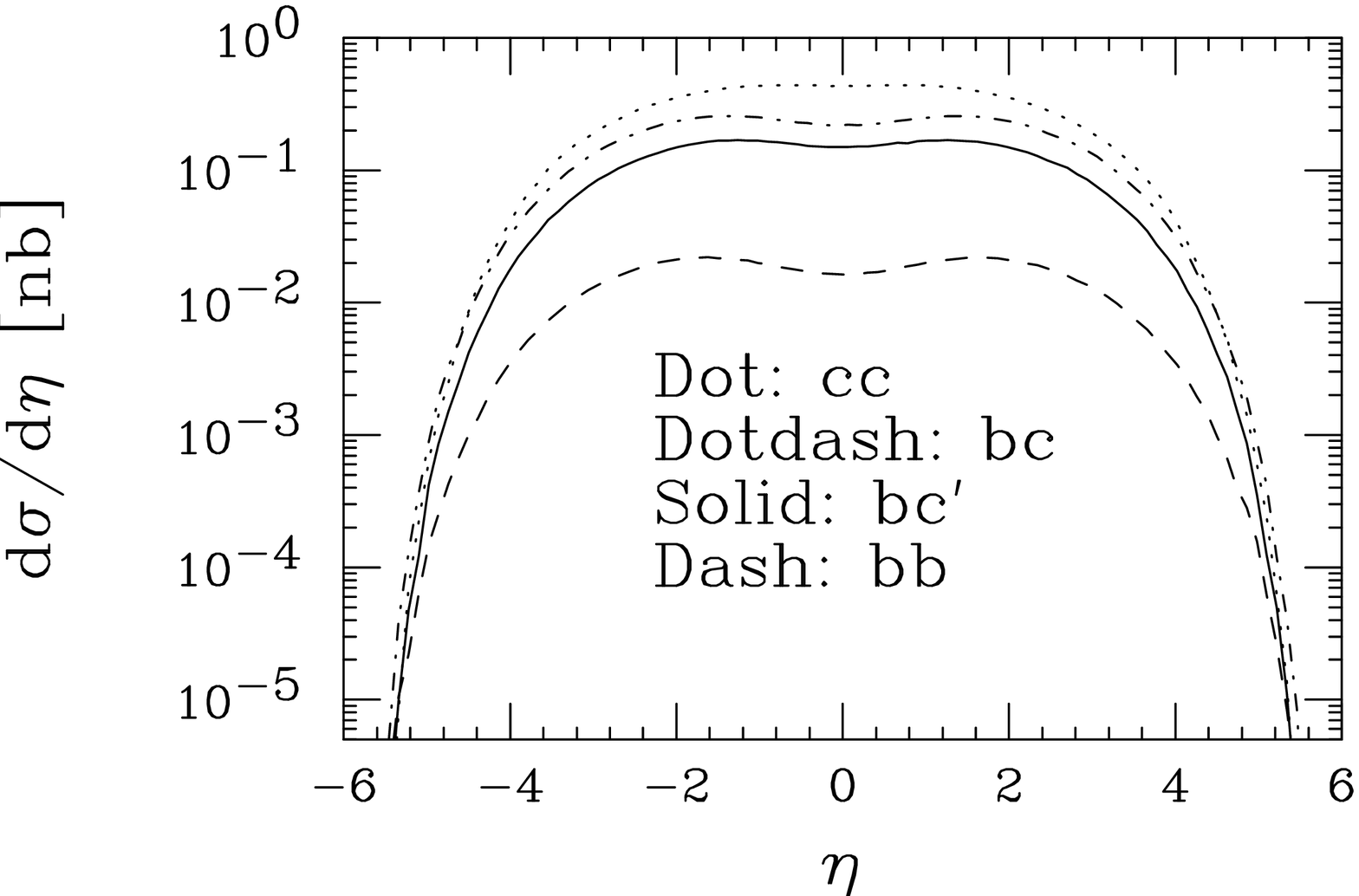,width=\linewidth}
\caption{Pseudorapidity distributions for doubly heavy baryon production at
the Tevatron.}
\label{fig:eta_tev}
\end{minipage}
\end{figure}

Some results are given in Figs.~\ref{fig:pt_tev} and \ref{fig:eta_tev}.
Fig.~\ref{fig:pt_tev} shows $d\sigma/dp_{_T}/d\eta|_{\eta=0}$ {\it vs.}
$p_{_T}$.  A kinematical cut of 5 GeV for the baryon $p_{_T}$ has been
imposed here.  Fig.~\ref{fig:eta_tev} gives the rapidity distribution of
$Q_1Q_2$ baryons, again with $p_{_T}>5$ GeV.  The rapidity distributions are
rather flat in the central region, so that $d\sigma/dp_{_T}$ may be
estimated from $d\sigma/dp_{_T}/d\eta|_{\eta=0}$ by multiplying with the
$\eta$-region desired.

% figures 1 and 2 --- Tevatron results
\noindent

% figures 3 and 4 --- LHC!
\noindent
\begin{figure}[hb]
\begin{minipage}{0.49\linewidth}
\epsfig{file=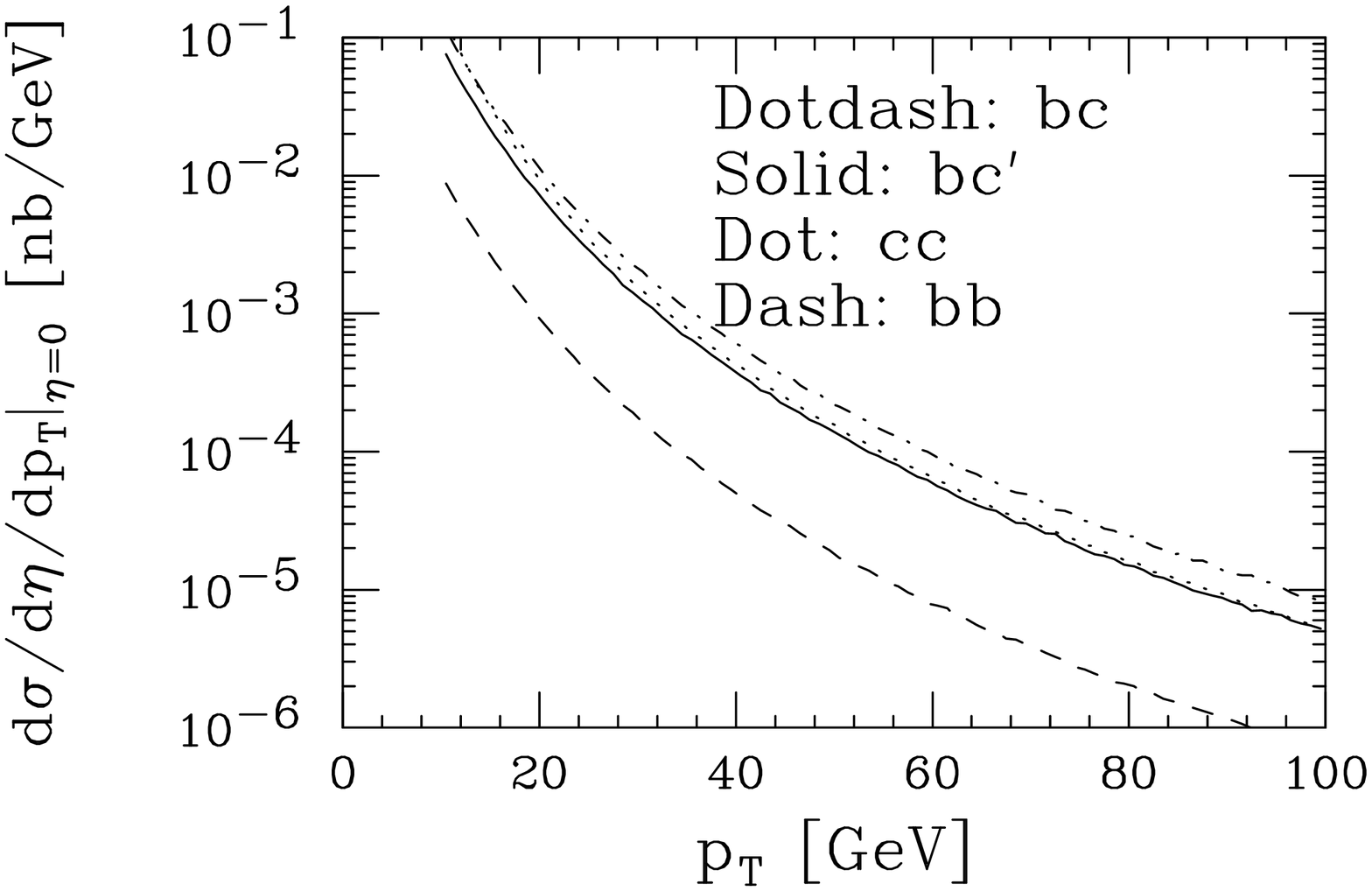,width=\linewidth}
\caption{$p_{_T}$ distributions (at small rapidity) for doubly heavy baryon
production at the LHC, 14 TeV.}
\label{fig:pt_lhc14}
\end{minipage}\hfill
\begin{minipage}{0.49\linewidth}
\epsfig{file=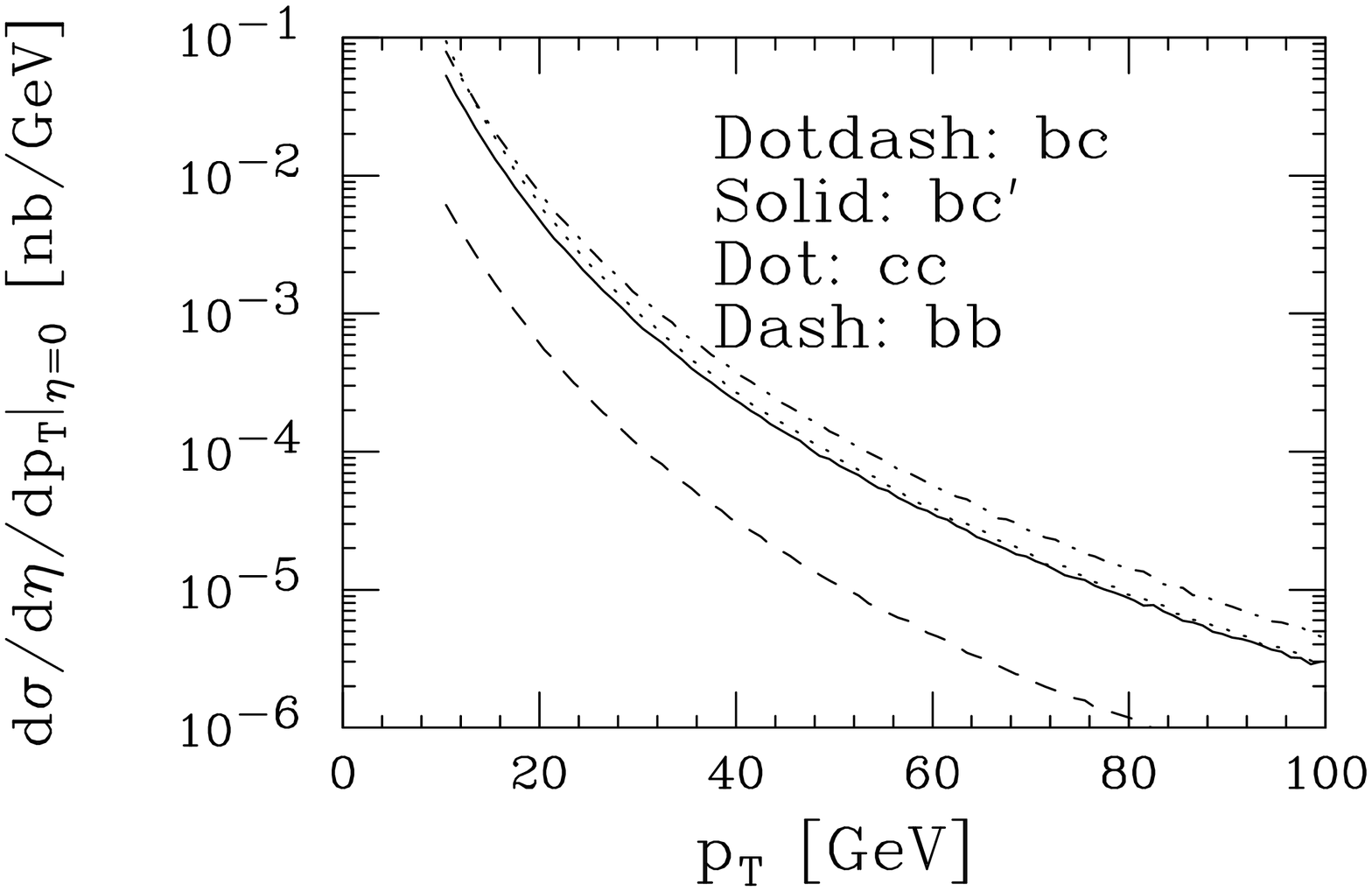,width=\linewidth}
\caption{$p_{_T}$ distributions (at small rapidity) for doubly heavy baryon
production at the LHC, 10 TeV.}
\label{fig:pt_lhc10}
\end{minipage}
\end{figure}

Figs.~\ref{fig:pt_lhc14} and \ref{fig:pt_lhc10} show
$d\sigma/dp_{_T}/d\eta|_{\eta=0}$ {\it vs.} $p_{_T}$ for the LHC operating
at 14~TeV (Fig.~\ref{fig:pt_lhc14}) and 10~TeV (Fig.~\ref{fig:pt_lhc10}).  A
kinematical cut of $p_{_T} > 10$ GeV is placed on the $Q_1Q_2$ baryon.  The
production rate at the lower center-of-mass energy is slightly lower, but
even at the lower energy, the LHC will produce $Q_1Q_2$ baryons copiously.
The rapidity distributions are again relatively flat in the central region
($|\eta|<3$), and event rates for any specific detector can be estimated by
simply multiplying the results in Table~\ref{tab:tot_sig} by the $\eta$
coverage of the detector.

\begin{figure}[h]
\begin{minipage}{0.49\linewidth}
\epsfig{file=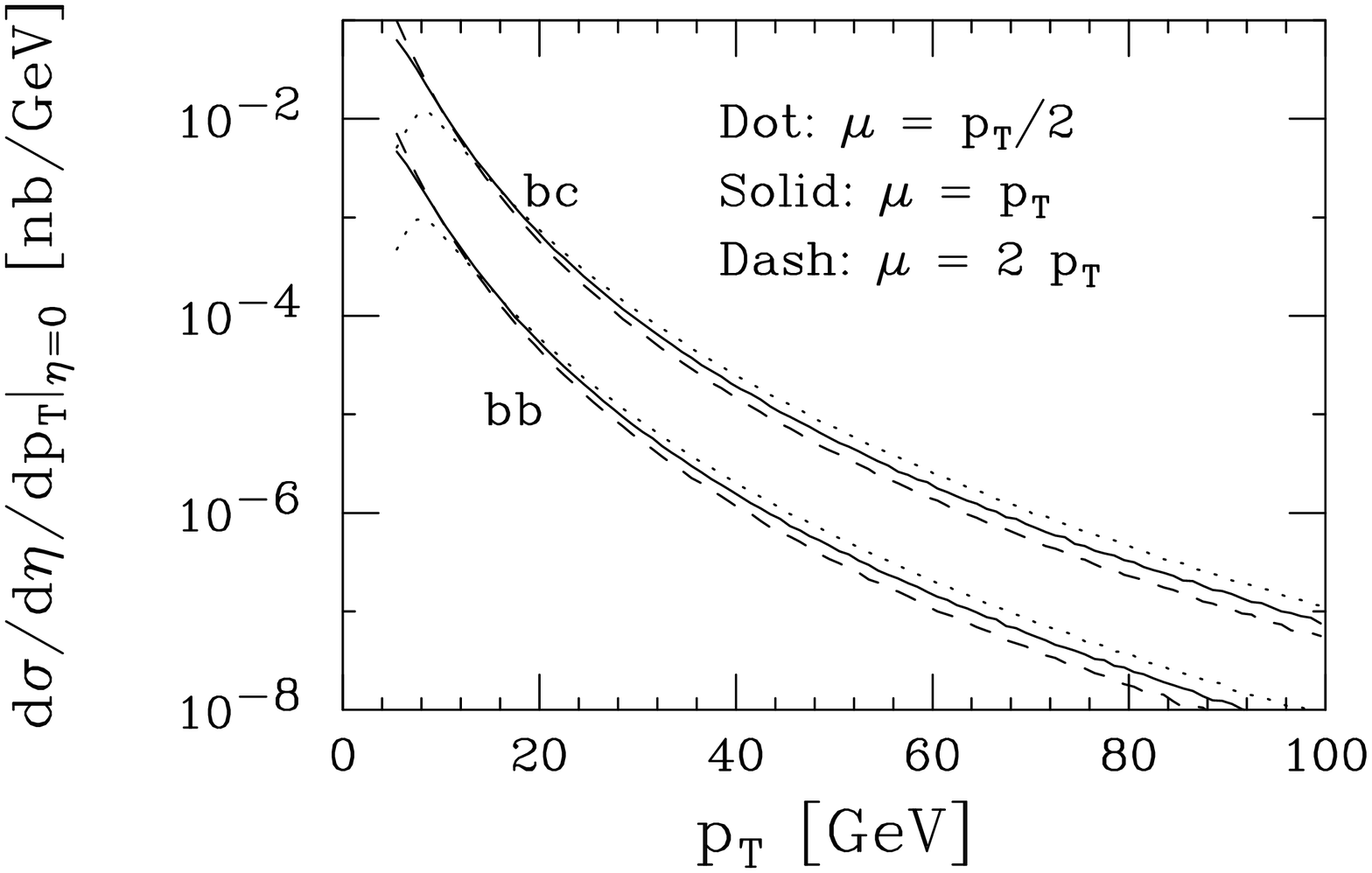,width=\linewidth}
\caption{$p_{_T}$ distributions (at small rapidity) for $bb$ and $bc$
production at the Tevatron for different choices of scale $\mu$.}
\label{fig:tev_pt_mu}
\end{minipage}\hfill
\begin{minipage}{0.49\linewidth}
\epsfig{file=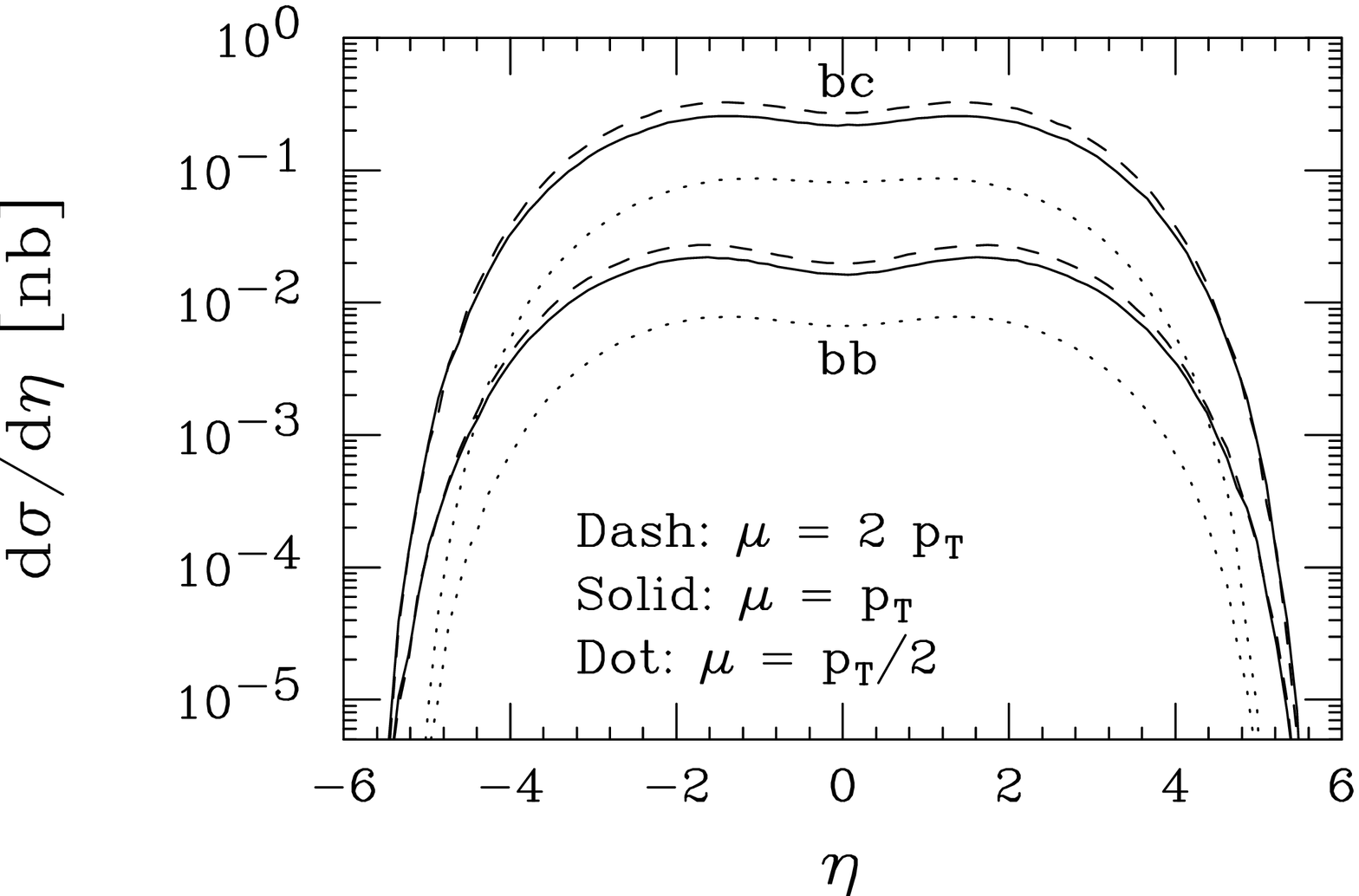,width=\linewidth}
\caption{Pseudorapidity distributions for $bb$ and $bc$ production at the
Tevatron for different choices of scale $\mu$.}
\label{fig:tev_eta_mu}
\end{minipage}
\end{figure}

In order to estimate the uncertainties due to the choice of factorization
and fragmentation scales, we vary these scales (all chosen to be equal) by a
factor of 2 from our nominal choice $\mu$ = $p_{_T}$.  The effect of varying
the scale on $d\sigma/dp_{_T}/d\eta|_{\eta=0}$ {\it vs.} $p_{_T}$ can be
seen in Fig.~\ref{fig:tev_pt_mu}, and the effect on $d\sigma/d\eta$
{\it vs.} $\eta$ can be seen in Fig.~\ref{fig:tev_eta_mu}.  The processes
shown are representative of all the processes.  The behavior seen in
Fig.~\ref{fig:tev_eta_mu} is somewhat counter-intuitive in that one expects
the rate with $\mu = p_{_T}/2$ to be larger than that for $\mu = p_{_T}$.
However, the fragmentation functions are cut off at low $\mu_{frag}$.  This
causes the drop at low $p_{_T}$ for this choice seen in
Fig.~\ref{fig:tev_pt_mu} and the lower cross sections in
Fig.~\ref{fig:tev_eta_mu}.  We use MRSA\cite{mrsa} parton distribution
functions.  The effect of changing the parton distribution set choice is
negligible --- far smaller than the effect of varying the choice of $\mu$.
Other sources of theoretical uncertainty include the parameters used for the
unevolved fragmentation functions, primarily $|R(0)|^2$ where different
phenomenological potentials give about 10\% differences in the result,
unknown QCD corrections to the parton-level cross sections, the
fragmentation function and the evolution equations, and unknown relativistic
corrections to the initial fragmentation functions.

% figures 5 and 6 --- TeV, mu dependence
\noindent

Recently there has been some concern that the fragmentation approximation
may not work as well as anticipated for $B_c$ production.  This assertion is
based on a comparison between the full calculation and the fragmentation
approximation for $B_c$ in $\gamma\gamma$ collisions\cite{kol1}.  There is,
of course, a significant difference between $\gamma\gamma$ and $gg$ initial
states, and the convolution of the gluon distributions with the parton-level
subprocess can have a large effect.  The calculation of $gg \to B_c X$ is
much more complex than that of $\gamma \gamma \to B_c X$, but is necessary
to understand the validity of the fragmentation approximation.  The authors
of \cite{kol1} went on to study $B_c$ production at hadron
colliders\cite{kol2}.  They found that the fragmentation approximation still
differs from the full calculation at the parton level
($\hat{\sigma}[g g \to b \bar{b}(\to B_c X)]$ {\it vs.}
$\hat{\sigma}[g g \to B_c b \bar{c}]$) except at high $p_{_T}$, but after
the convolution of $\hat{\sigma}$ with the gluon distribution functions, the
full calculation and the fragmentation approximation agree quite well for
$p_{_T} > 10$~GeV both at Tevatron and LHC energies.  Fig.~6 of \cite{kol2}
shows that the agreement between the full calculation and the fragmentation
approximation is acceptable down to $p_{_T}^{min} = 5$~GeV at the Tevatron
and $p_{_T}^{min} = 10$~GeV at LHC.  As the fragmentation functions and the
production mechanisms are extremely similar for $B_c$ and heavy diquark
production, our calculation of the production of doubly heavy baryons should
also be accurate in this $p_{_T}$ range.

The decay modes of doubly charmed baryons have been studied in \cite{rox}
using SU(3) flavor symmetry.  This approach relates the decays of these
particles, and could be used to find relations among the decays of the $bb$
baryons and among the $bc$ baryons.  The transitions between the doubly
heavy baryons\cite{whitesavage} and some decays of the $bc$
baryons\cite{sanchis} have been discussed in the heavy quark limit.  The
lifetimes of $c$ and $b$ quarks are not very different
($\tau_{_{\Lambda_c}} \sim 0.2 \times 10^{-12}$~s while
$\tau_{_{\Lambda_b}} \sim 1.07 \times 10^{-12}$~s\cite{pdg}), so that the
weak decays of both $b$ and $c$ are important when considering the decays of
the $bc$ baryons.  $bc$ baryons will decay weakly (both hadronically and
semi-leptonically) to single $b$ quark baryons as well as doubly charmed
baryons.  Both of these possibilities have been discussed in the
literature\cite{rox,fleckrich,datta}.  Either a single $b$ baryon or a
doubly charmed baryon traced back to a displaced vertex using a vertex
detector will be a clear sign of doubly heavy baryon production\cite{baur}.
In the case of $bb$ baryons, the production of a doubly charmed baryon
(possibly with same sign di-lepton!), with both $b$-quark decays tagged by a
vertex detector, would be an impressive signal.  Unfortunately, the $bb$
baryons are produced less frequently than the other doubly heavy baryons,
and two semi-leptonic decays will severely reduce the event rate.

In conclusion, the production via fragmentation of baryons containing two
heavy quarks has been calculated in the fragmentation approximation.  The
fragmentation approximation reproduces well the full calculation of $B_c$
production at hadron colliders for the $p_{_T}$ studied here\cite{kol2}; it
is expected that the fragmentation approximation will work well for doubly
heavy diquark production as well.  The production rates of these baryons at
$e^+e^-$ colliders and at the $ep$ collider HERA are found to be
negligible.  The situation is much better at hadron colliders, with
approximately $8\times10^4$ events/yr expected at Tevatron and
$1.3\times10^8$ events/year at LHC.  In addition to predictions for total
production cross sections ($|\eta|<0.5$ and $p_{_T}>5(10)$~GeV at the
Tevatron (LHC)), the distributions $d\sigma/dp_{_T}/d\eta|_{\eta=0}$ and
$d\sigma/d\eta$ for $p_{_T}>5(10)$~GeV at the Tevatron (LHC) are studied.
The detection of these particles and measurements of their properties will
provide an experimental challenge due to the large number of their decay
modes, but will be a rich testing ground, {\it e.g.}, for HQET and QCD
potential models.

\vspace*{0.4cm}
\noindent{\Large {\bf Acknowledgments}} \\
The work of M.A.D. was supported by the Natural Sciences and Engineering
Research Council (NSERC) --- Canada.

\end{document}